\documentclass[showpacs,amsmath,pra]{revtex4}
\usepackage{graphicx}
\usepackage{dcolumn}
\usepackage{bm}
\begin{document}

\title{K-dimensional trio coherent states}
\author{\textbf{Hyo Seok Yi}}
\email{hsyi@muon.kaist.ac.kr}
\affiliation{Department of Physics, Korea Advanced Institute of
Science and Technology, 373-1 Guseong-dong, Yuseong-gu, Daejeon 305-701,
Republic of Korea}
\author{\textbf{Ba An Nguyen}}
\email{nbaan@kias.re.kr}
\thanks{Corresponding author}
\affiliation{School of Computational Sciences, Korea Institute for Advanced
Study, 207-43 Cheongryangni 2-dong, Dongdaemun-gu, Seoul 130-722, Republic
of Korea}
\author{\textbf{Jaewan Kim}}
\email{jaewan@kias.re.kr}
\affiliation{School of Computational Sciences, Korea Institute for Advanced
Study, 207-43 Cheongryangni 2-dong, Dongdaemun-gu, Seoul 130-722, Republic
of Korea}

\begin{abstract}
We introduce a novel class of higher-order, three-mode states called
K-dimensional trio coherent states. We study their mathematical properties
and prove that they form a complete set in a truncated Fock space. We also
study their physical content by explicitly showing that they exhibit
nonclassical features such as oscillatory number distribution,
sub-poissonian statistics, Cauchy-Schwarz inequality violation and
phase-space quantum interferences. Finally, we propose an experimental
scheme to realize the state with $K=2$ in the quantized vibronic motion of a
trapped ion.
\end{abstract}

\maketitle

\noindent \textbf{1. Introduction}

\noindent Higher-order effect as well as multimode nature of quantum states
play a significant role in various physical applications. For instance,
while multimode entangled states serve as necessary resources in multiuser
quantum communication network (see, e.g., \cite{tele1,tele2,t3}),
higher-order character gives rise to simultaneous squeezing in several
directions \cite{ho1,ho2,ho3} or to enhancement of antibunching \cite{enh}, etc.
Here we deal with states that are higher-order and three-mode. 

The motivation for studying such states is twofold. Firstly, it is the academic motivation 
since we shall go on a route similar to those from $1$-dimensional single-mode (two-mode) coherent states 
to $K$-dimensional single-mode \cite{sing1,sing2,sing3,sing4,sing5,sing6,sing7} 
(two-mode \cite{two1,two2,two3,two4}) coherent states. Our study concerns the case 
of three modes thus generalizing the previously existing works. Our three-mode states 
are intrinsically nonclassical (see \cite{do} for a comprehensive review). 
It is worth mentioning here that a number of nonclassical states 
are intimately linked with a symmetry group. But this does not compulsorily hold for all nonclassical states. 
For example, the $N$-mode sum-squeezed states \cite{sum1,sum2} are always (i.e., for any $N\ge 2$) 
connected with the su(1,1) Lie algebra. Yet, the operators characterizing the $N$-mode 
difference-squeezed state \cite{sum1,diff1} forms a representation of the su(2) Lie algebra only for $N=2$,  
not for $N\ge 3$  \cite{diff2}. The states we are going to study in this paper have no relation to a symmetry group. 
Secondly, the practical motivation could also be foreseen. As will be shown, 
a $K$-dimensional state is in fact a quantum superposition of $K$ massive distinguishable substates. 
Such superpositions not only help to deeper understand the physics world both from quantum and classical point 
of view, but they may also serve as resources for quantum computation and quantum information processing. 
Moreover, our three-mode states are {\it naturally} 
entangled and, as such, they potentially promise wide applications.  
In this connection we notice that pair coherent states (PCS) \cite{pair}, 
which are of two-mode entangled character,   
have been found useful in testing fundamental concepts of physics \cite{r94,r98,r99,r04} as well as in accomplishing 
quantum teleportation \cite{t03,t04}.  We therefore do believe that there should be 
tasks (say,  with participation of three parties like Fredkin or Toffoli gates and so on)  which cannot be performed by 
single- or two-mode states but can be performed by three-mode ones such as ours. 
 
The states of interest are defined as the eigenstates of the operator $(\widehat{a}%
\widehat{b}\widehat{c})^{K},$ with $K$ a positive integer and $\widehat{a}$ $%
(\widehat{b},$ $\widehat{c})$ the boson annihilation operator of mode $a$ $%
(b,c),$ subject to the conditions that they be also the eigenstates of the
operators $\widehat{P}=\widehat{b}^{\dagger}\widehat{b}-\widehat{c}^{\dagger}\widehat{c}$
and $\widehat{Q}=\widehat{a}^{\dagger}\widehat{a}-\widehat{c}^{\dagger}\widehat{c}$
corresponding to the differences of the number of quanta in the three modes. It is noted that 
the operators $\widehat{a}$,  $\widehat{b}$ and  $\widehat{c}$ commute between each other. 
Denoting the state by $\left| \xi ,p,q\right\rangle _{K},$ the three following
equations must be simultaneously satisfied

\begin{equation}
(\widehat{a}\widehat{b}\widehat{c})^{K}\left| \xi ,p,q\right\rangle _{K}=\xi
^{K}\left| \xi ,p,q\right\rangle _{K}\text{ },  \label{k}
\end{equation}
\begin{equation}
\widehat{P}\left| \xi ,p,q\right\rangle _{K}=p\left| \xi ,p,q\right\rangle
_{K}\text{ },  \label{p}
\end{equation}
\begin{equation}
\widehat{Q}\left| \xi ,p,q\right\rangle _{K}=q\left| \xi ,p,q\right\rangle
_{K}\text{ }  \label{q}
\end{equation}
where $\xi =r\exp (i\varphi )\in \mathcal{C}$ and $p,q$ are referred to as
``charges'' \cite{charge1,charge2} which, without loss of generality, are
assumed to be non-negative. Equations (\ref{k}) to (\ref{q}) signify that a
superposition of three-mode Fock states in the form

\begin{equation}
\left| \xi ,p,q\right\rangle _{K}=\sum_{n=0}^{\infty }c_{n}\left|
n+q\right\rangle _{a}\left| n+p\right\rangle _{b}\left| n\right\rangle _{c}
\label{f}
\end{equation}
is a solution provided the coefficients $c_{n}$ obey the constraint

\begin{equation}
c_{n+K}=\sqrt{\frac{(n+q)!(n+p)!n!}{(n+q+K)!(n+p+K)!(n+K)!}}\,\,\xi ^{K}c_n 
\text{.}  \label{cn}
\end{equation}
This constraint forms $K$ independent strings $\{s_{0},s_{1},...,s_{K-1}\}$
where each of $s_{j}$ consists of an infinite number of related coefficients 
$s_j=\{c_{j},c_{j+K},c_{j+2K},...\}$ with $c_{j}$ being the representative
of the $s_{j}$-string. The states $\left| \xi ,p,q\right\rangle _{K}$ are
therefore highly degenerate since the representatives $c_{0},$ $c_{1},$ ..., 
$c_{K-1}$ remain arbitrary. We now impose conditions on the representative $%
c_{j}$ so that a state $\left| \xi ,p,q\right\rangle _{Kj}$ associated with
the coefficients taken from the $s_{j}$-string be normalized to unity. (Note that 
the subscript $Kj$ should not be conceived as a single index; it refers to two separate 
integers $K$ and $j$.) Then,
each state $\left| \xi ,p,q\right\rangle _{Kj}$ with a fixed $j\in [0,K-1]$
has the following explicit Fock expansion

\begin{equation}
\left| \xi ,p,q\right\rangle _{Kj}=N_{Kj}(r^{2},p,q)\sum_{n=0}^{\infty }%
\frac{\xi ^{Kn+j}}{\sqrt{\rho _{pq0}(Kn+j)}}\left| Kn+j+q\right\rangle
_{a}\left| Kn+j+p\right\rangle _{b}\left| Kn+j\right\rangle _{c}  \label{fs}
\end{equation}
where 
\begin{equation}
\rho _{pq0}(Kn+j)=(Kn+j+p)!(Kn+j+q)!(Kn+j)!  \label{roKj}
\end{equation}
and

\begin{equation}
N_{Kj}(x,p,q)=\left( \sum_{m=0}^{\infty }\frac{x^{Km+j}}{\rho _{pq0}(Km+j)}%
\right) ^{-1/2}  \label{C}
\end{equation}
is the normalization coefficient. Since the overlap between states $\left|
\xi ,p,q\right\rangle _{Kj}$ and $\left| \xi ^{\prime },p^{\prime
},q^{\prime }\right\rangle _{Kj^{\prime }}$ is determined by

\begin{equation}
_{Kj^{\prime }}\left\langle \xi ^{\prime },p^{\prime },q^{\prime }\right.
\left| \xi ,p,q\right\rangle _{Kj}=\delta _{jj^{\prime }}\delta _{pp^{\prime
}}\delta _{qq^{\prime }}\frac{N_{Kj}(\xi ^{\prime *},p,q)N_{Kj}(\xi ,p,q)}{%
N_{Kj}^{2}(\sqrt{\xi ^{\prime *}\xi },p,q)}  \label{scalar}
\end{equation}
it follows that the normalized states $\left| \xi ,p,q\right\rangle _{Kj}$
are orthogonal with respect to $j,p,q$ but non-orthogonal with respect to $%
\xi .$ It is worth emphasizing that for a given $K$ there are $K$ states $%
\left| \xi ,p,q\right\rangle _{Kj}$ $(j=0,1,...,K-1)$ each of which is a
three-mode entangled state. In other words, the eigenvalue set $\{\xi
^{K},p,q\}$ is $K$-degenerate corresponding to $K$ linearly independent
eigenstates $\left| \xi ,p,q\right\rangle _{Kj}.$ We call them K-dimensional
trio coherent states (KTCS's) because the dimension of the space spanned by
these states is $K$ even though each of them is embedded in a vector space
of infinite dimension characterized by the complex variable $\xi .$ In the
same spirit, multidimensional entangled coherent states have also been
investigated recently in \cite{mecs} where they prove to be necessary for
teleportation of quantum states of a particular kind. When $K=1$ the KTCS
reduces to the trio coherent state (TCS) introduced in \cite{tcs} which is a
generalization of the PCS \cite{pair}. So, in what
follows, when KTCS's are quoted it implies $K>1.$

In the next part of this paper we first expose mathematical properties of
the KTCS including proof of their (over)completeness. Physical properties
are then studied showing that KTCS's are inherently nonclassical. Finally,
we propose an experimental scheme to produce a KTCS in the vibrational
motion of the center-of-mass of an ion trapped in real space by a
three-dimensional harmonic potential. \vskip 0.5cm

\noindent \textbf{2. Mathematical properties}

\noindent States $\left| \xi ,p,q\right\rangle _{Kj}$ with different $j$ can
be transformed from one to another by a ``rotation'' operator $\widehat{R}%
_{Klm}(\xi ,p,q)$ defined by 
\begin{equation}
\widehat{R}_{Klm}(\xi ,p,q)=\frac{N_{Kl}(r^{2},p,q)}{N_{Km}(r^{2},p,q)}\xi
^{-[m-l]_{K}}(\widehat{a}\widehat{b}\widehat{c})^{[m-l]_{K}}  \label{R}
\end{equation}
where $[x]_{K}=x$ if $x\geq 0 $ and $[x]_{K}=x+K$ if $x<0.$ That is, for any 
$l,m\in [0,K-1],$

\begin{equation}
\left| \xi ,p,q\right\rangle _{Kl}=\widehat{R}_{Klm}(\xi ,p,q)\left| \xi
,p,q\right\rangle _{Km}.  \label{lm}
\end{equation}

Another important property is: any KTCS can be expressed as a superposition
of $K$ phase-correlated TCS's. Namely,

\begin{equation}
\left| \xi ,p,q\right\rangle _{Kj}=\frac{N_{Kj}(r^{2},p,q)}{KN(r^{2},p,q)}%
\sum_{j^{\prime }=0}^{K-1}\exp \left( -\frac{2\pi i}{K}jj^{\prime }\right)
\left| \xi _{Kj^{\prime }},p,q\right\rangle  \label{Kto1}
\end{equation}
where $\left| \xi ,p,q\right\rangle \equiv \left| \xi ,p,q\right\rangle
_{10},$ $N(r^{2},p,q)\equiv N_{10}(r^{2},p,q)$ and

\begin{equation}
\xi _{Kj}=\xi \exp \left( \frac{2\pi ij}{K}\right) .  \label{xiKj}
\end{equation}
The expansion (\ref{Kto1}) can be straightforwardly verified by virtue of
the identities $\sum_{m=0}^{K-1}\exp \left[ 2\pi i(l-l^{\prime })m/K\right]
\equiv K\delta _{ll^{\prime }}$ and $\sum_{m=0}^{\infty }X_{m}\left|
m\right\rangle \equiv \sum_{l=0}^{K-1}\sum_{n=0}^{\infty }X_{nK+l}\left|
nK+l\right\rangle .$ It can be noted, due to Eq. (\ref{xiKj}), that the
TCS's superposing a KTCS are evenly distributed on a circle of radius $r$ in
the phase space. The inverse transformation of (\ref{Kto1}) is

\begin{equation}
\left| \xi _{Kj},p,q\right\rangle =N(r^{2},p,q)\sum_{j^{\prime }=0}^{K-1}%
\frac{\exp \left( \frac{2\pi i}{K}jj^{\prime }\right) }{N_{Kj^{\prime
}}(r^{2},p,q)}\left| \xi ,p,q\right\rangle _{Kj^{\prime }}.  \label{1toK}
\end{equation}
Alternatively, making a change $\{\xi _{Kj}\equiv \xi \exp (2\pi
ij/K)\rightarrow \chi ,$ $\xi \rightarrow \chi \exp (-2\pi ij/K)\}$ in (\ref
{1toK}) and taking into account the identity

\begin{equation}
\left| \chi \exp (-2\pi ij/K),p,q\right\rangle _{Kj^{\prime }}\equiv \exp
(-2\pi ijj^{\prime }/K)\left| \chi ,p,q\right\rangle _{Kj^{\prime }}
\label{KK}
\end{equation}
which can be easily checked by use of the Fock representation (\ref{fs}) in
both sides of (\ref{KK}), we can cast (\ref{1toK}) into a simpler but more
convenient form (after changing $\chi $ back to $\xi )$ as

\begin{equation}
\left| \xi ,p,q\right\rangle =N(r^{2},p,q)\sum_{j=0}^{K-1}\frac{\left| \xi
,p,q\right\rangle _{Kj}}{N_{Kj}(r^{2},p,q)}.  \label{Also1toK}
\end{equation}
More interestingly, it turns out that KTCS's of any two different dimensions
are also related to each other. A hint to establish such a relation is first
using (\ref{Kto1}) to express a KTCS with a given $K$ in terms of TCS's and
then applying (\ref{Also1toK}) to get the TCS's back in terms of K$'$TCS's
with a different $K^{\prime }$ (generally $K^{\prime }\neq K).$ As a result,
we arrive at

\begin{equation}
\left| \xi ,p,q\right\rangle _{Kj}=\frac{N_{Kj}(x,p,q)}{K}\sum_{j^{\prime
}=0}^{K^{\prime }-1}\sum_{j^{\prime \prime }=0}^{K-1}\frac{\exp \left( -%
\frac{2\pi i}{K}jj^{\prime \prime }\right) }{N_{K^{\prime }j^{\prime
}}(x,p,q)}\left| \xi _{Kj^{\prime \prime }},p,q\right\rangle _{K^{\prime
}j^{\prime }}.  \label{KtoK}
\end{equation}
The transformation (\ref{KtoK}) is most general in the sense that it
recovers (\ref{Kto1}) when $K^{\prime }=1,j^{\prime }=0$ and (\ref{Also1toK}%
) when $K=1,j=0.$ In the special case, when $K^{\prime }=K,$ the r.h.s. of (%
\ref{KtoK}) becomes nothing else but its l.h.s., as it should be.

In terms of the usual single-mode coherent state

\begin{equation}
|\eta )\equiv \exp (-|\eta |^{2}/2)\sum_{n=0}^{\infty }\frac{\eta ^{n}}{%
\sqrt{n!}}\left| n\right\rangle ,  \label{ucs}
\end{equation}
the following expansion holds in general

\begin{eqnarray}
\left| \xi ,p,q\right\rangle _{Kj} &=&N_{Kj}(r^{2},p,q)\int \frac{\alpha
^{*q}d^{2}\alpha }{\pi }\int \frac{\beta ^{*p}d^{2}\beta }{\pi }\int \frac{%
d^{2}\gamma }{\pi }N_{Kj}^{-2}(\xi \alpha ^{*}\beta ^{*}\gamma ^{*},p,q) 
\nonumber \\
&&\times \exp \left[ -\left( |\alpha |^{2}+|\beta |^{2}+|\gamma |^{2}\right)
/2\right] |\alpha )_{a}|\beta )_{b}|\gamma )_{c}  \label{expcs}
\end{eqnarray}
because of the completeness of the coherent state. In particular, however, a
formula via three phase-correlated coherent states whose amplitudes $\alpha
, $ $\beta $ and $\gamma $ satisfy the equality $\alpha \beta \gamma =\xi $
seems to be more useful. That looks like this 
\begin{eqnarray}
\left| \xi ,p,q\right\rangle _{Kj} &=&\frac{1}{K}N_{Kj}(r^{2},p,q)\exp
\left[ \left( |\alpha |^{2}+|\beta |^{2}+|\gamma |^{2}\right) /2\right] 
\nonumber \\
&&\times \sum_{j^{\prime }=0}^{K-1}\frac{\exp \left( -\frac{2\pi i}{K}%
jj^{\prime }\right) }{\alpha _{Kj^{\prime }}^{q}\beta _{Kj^{\prime }}^{p}}%
\int_{0}^{2\pi }\frac{d\theta }{2\pi }\int_{0}^{2\pi }\frac{d\theta ^{\prime
}}{2\pi }\exp [-i(q\theta +p\theta ^{\prime })]  \nonumber \\
&&\times \left| \alpha _{Kj^{\prime }}\exp (i\theta )\right) _{a}\left|
\beta _{Kj^{\prime }}\exp (i\theta ^{\prime })\right) _{b}\left| \gamma
_{Kj^{\prime }}\exp [-i(\theta +\theta ^{\prime })]\right) _{c}.  \label{cs}
\end{eqnarray}
In fact, when $K=1,$ $j=0$ and $\alpha =\beta =\gamma =\xi ^{1/3},$ formula (%
\ref{cs}) reduces to formula (1) in \cite{yi}.

We now turn to the important issue of proving that the $K$ states $\left|
\xi ,p,q\right\rangle _{Kj}$ $(j=0,1,...,K-1)$ form a complete set. As seen
from (\ref{fs}), any state $\left| \xi ,p,q\right\rangle _{Kj}$ spans the
three-mode Fock space in which the $q$ $(p)$ first number states of mode $a$ 
$(b)$ with $n_{a}=0,1,...,q-1$ $(n_{b}=0,1,...,p-1)$ are lacking. In such a
truncated Fock space the unity operator has to be of the form (similar
matter associated with single-mode photon-added coherent states can be found
in \cite{pacs})

\begin{equation}
\mathbf{I}_{p,q}=\sum_{n=0}^{\infty }\left| n+q\right\rangle _{a}\left|
n+p\right\rangle _{b}\left| n\right\rangle _{c}\left\langle n\right| _{\text{
}b}\left\langle n+p\right| _{\text{ }a}\left\langle n+q\right| .  \label{I}
\end{equation}
The resolution of unity of the KTCS's then amounts to existence of a weight
function $W_{Kj}(|\xi |^{2},p,q)$ such as to fulfill the following
completeness condition

\begin{equation}
\sum_{j=0}^{K-1}\int d^{2}\xi \left| \xi ,p,q\right\rangle _{Kj}W_{Kj}(|\xi
|^{2},p,q)_{Kj}\left\langle \xi , p,q \right| =\mathbf{I}_{p,q}.
\label{completeness}
\end{equation}
To solve for $W_{Kj}(|\xi |^{2},p,q)$ we substitute $\xi =r\exp (i\varphi )$
and (\ref{fs}) into (\ref{completeness}) and integrate over $\varphi .$
After the $\varphi $-integration the function $W_{Kj}(r^{2},p,q)$ is looked
for in the form

\begin{equation}
W_{Kj}(r^{2},p,q)=\frac{\widetilde{W}(r^{2},p,q,0)}{\pi N_{Kj}^{2}(r^{2},p,q)%
},  \label{w}
\end{equation}
where $\widetilde{W}(r^{2},p,q,0)$ is to be determined from the equation

\begin{equation}
\int_{0}^{\infty }\widetilde{W}(x,p,q,0)x^{n}dx=\rho _{pq0}(n);\text{ }%
n=0,1,2,...,\infty .  \label{wnga}
\end{equation}
This is the classical Stieltjes power-moment problem \cite{moment}, with the
set of $n$-$th$ moments $\rho _{pq0}(n)$ parameterized by $\{p,q,0\}.$ If $n$
in Eq. (\ref{wnga}) is extended to $s-1$ where $s\in \mathcal{C},$ then our
problem can be formulated in terms of the Mellin and inverse Mellin
transforms \cite{mellin1,mellin2} which have been extensively used in the
context of various kinds of generalized coherent states \cite
{gcs1,gcs2,gcs3,gcs4,gcs5,gcs6}. Here, $\rho _{pq0}(s-1)$ is the Mellin
transform, $\mathcal{M}[\widetilde{W}(x,p,q,0);s],$ of $\widetilde{W}%
(x,p,q,0),$ i.e.

\begin{equation}
\rho _{pq0}(s-1)=\mathcal{M}[\widetilde{W}(x,p,q,0);s]:=\int_{0}^{\infty }%
\widetilde{W}(x,p,q,0)x^{s-1}dx
\end{equation}
and, $\widetilde{W}(x,p,q,0)$ in turn is the inverse Mellin transform, $%
\mathcal{M}^{-1}[\rho _{pq0}(s-1);x],$ of $\rho _{pq0}(s-1),$ i.e.

\begin{equation}
\widetilde{W}(x,p,q,0)=\mathcal{M}^{-1}[\rho _{pq0}(s-1);x] :=\frac{1}{2\pi i%
}\int_{-i\infty }^{i\infty}\rho _{pq0}(s-1)x^{-s}ds.  \label{wn2}
\end{equation}
We know \cite{gcs5} that the solution of the simpler problem

\begin{equation}
\int_{0}^{\infty }\widetilde{W}(x,l)x^{n}dx=\rho _{l}(n)=(n+l)!,  \label{rl}
\end{equation}
is given by 
\begin{equation}
\widetilde{W}(x,l)=\mathcal{M}^{-1}[\rho _{l}(s-1);x]=x^{l}e^{-x}.
\label{wn4}
\end{equation}
To make use of the known result (\ref{wn4}) we twice apply the Parseval
relation \cite{mellin1,mellin2} which we reformulate for our purpose here in
the form

\begin{equation}
\mathcal{M}^{-1}[f(s-1)g(s-1);x]=\int_{0}^{\infty }\frac{dt}{t}\mathcal{M}%
^{-1}[f(s-1);t]\mathcal{M}^{-1}[g(s-1);\frac{x}{t}].  \label{parseval}
\end{equation}
Now, by virtue of (\ref{wn2}), (\ref{rl}) and (\ref{parseval}), we have
immediately

\begin{eqnarray}
\widetilde{W}(x,p,q,0) &=&\mathcal{M}^{-1}[\rho _{pq0}(s-1);x]\equiv 
\mathcal{M}^{-1}[\rho _{p}(s-1)\rho _{q}(s-1)\rho _{0}(s-1);x]  \nonumber \\
&=&\int_{0}^{\infty }\frac{dt}{t}\mathcal{M}^{-1}[\rho _{p}(s-1)\rho
_{q}(s-1);t]\mathcal{M}^{-1}[\rho _{0}(s-1);\frac{x}{t}]  \nonumber \\
&=&\int_{0}^{\infty }\frac{dt}{t}\left[ \int_{0}^{\infty }\frac{d\tau }{\tau 
}\mathcal{M}^{-1}[\rho _{p}(s-1);\tau ]\mathcal{M}^{-1}[\rho _{q}(s-1);\frac{%
t}{\tau }]\right] \mathcal{M}^{-1}[\rho _{0}(s-1);\frac{x}{t}]  \nonumber \\
&=&\int_{0}^{\infty }t^{q-1}e^{-x/t}\left[ \int_{0}^{\infty }\tau
^{p-q-1}e^{-\tau -t/\tau }d\tau \right] dt  \label{wn5}
\end{eqnarray}
where in the last step we made use of Eq. (\ref{wn4}). Performing the $\tau $%
-integration we finally arrive at

\begin{equation}
\widetilde{W}(x;p,q,0)=\int_{0}^{\infty }t^{-1+(p+q)/2}e^{-x/t}K_{q-p}(2%
\sqrt{t})dt  \label{wngafinal}
\end{equation}
where $K_{n}(t)=K_{-n}(t)=K_{|n|}(t)$ stands for the modified Bessel
function of the second kind. There are two remarks to be made at this point
as follows.

\begin{itemize}
\item  The condition (\ref{completeness}) is that of over-completeness
rather than completeness. This is due to the non-orthogonality of KTCS's
with respect to $\xi $ and is explicitly reflected by the fact that any
state $\left| \xi ,p,q\right\rangle _{Kj}$ can be represented via the others
as 
\begin{equation}
\left| \xi ,p,q\right\rangle _{Kj}=N_{Kj}(\xi ,p,q)\int d^{2}\xi ^{\prime }%
\frac{N_{Kj}(\xi ^{\prime *},p,q)W_{Kj}(|\xi ^{\prime }|^{2},p,q)}{%
N_{Kj}^{2}(\sqrt{\xi ^{\prime *}\xi },p,q)}\left| \xi ^{\prime
},p,q\right\rangle _{Kj}.
\end{equation}

\item  The solution (\ref{wngafinal}) is not unique. According to the
Carleman's (sufficient) condition \cite{car}, our solution would be unique
(non-unique) if the sum $S=\sum_{n=1}^{\infty }S_{n},$ with $S_{n}=\left[
(Kn+j+q)!(Kn+j+p)!(Kn+j)!\right] ^{-1/2n},$ diverges (converges). We now
apply the logarithmic test \cite{test} to examine the convergence of $S.$
The logarithmic criterion says that if $T=\lim_{n\rightarrow \infty }\ln
(S_{n})/\ln (n)>-1$ $(<-1)$ then $S$ diverges (converges). To calculate $T$
we rewrite $S_{n}$ in terms of Gamma functions, $S_{n}=\left[ \Gamma
(Kn+j+q+1)\Gamma (Kn+j+p+1)\Gamma (Kn+j+1)\right] ^{-1/2n},$ and then use
the Stirling's formula, $\Gamma (Kn+l)\approx \sqrt{2\pi }\exp
(-Kn)(Kn)^{Kn},$ which is valid for large $n$ and finite $l$ $(l\ll n).$ As
a result, we obtain $\lim_{n\rightarrow \infty }\ln (S_{n})=-3K\ln (n)/2$
and, hence, $T=-3K/2<-1$ for any positive integers $K.$ This proves the
non-uniqueness of our solution (\ref{wngafinal}).
\end{itemize}

To end this section we remind from (\ref{fs}) that the KTCS's are normalized
by the normalization coefficient $N_{Kj}(r^{2},p,q)$ determined by (\ref{C}%
). The states are also continuous in the variable $\xi $ because $\left|
\left| \xi ,p,q\right\rangle _{Kj}-\left| \xi ^{\prime },p,q\right\rangle
_{Kj}\right| ^{2}= 2\left[ 1-\mathop{\rm Re}\left( _{Kj}\left\langle \xi
^{\prime },p,q \right. \left| \xi ,p,q\right\rangle _{Kj}\right) \right] $
tends to vanish when $\left| \xi -\xi ^{\prime }\right| \rightarrow 0,$ as
is evident from (\ref{C}) and (\ref{scalar}). Furthermore, as followed from (%
\ref{w}) and (\ref{wngafinal}), the weight function $W_{Kj}(|\xi |^{2},p,q)$
in (\ref{completeness}) is guaranteed to be positive (thanks to positivity
of the Bessel function $K_n(t)$). Therefore, the minimal set of conditions 
\cite{cs1,cs2} required for an ensemble of states to be \textit{coherent} is
met for the KTCS's. That explains why \textit{``coherent"} arises in naming
the states $\left| \xi ,p,q\right\rangle _{Kj}$ as \textit{``K-dimensional
trio coherent states''} of which \textit{``coherent''} must be family name
and \textit{``K-dimensional trio''} first name. \vskip 0.5cm

\noindent \textbf{3. Physical content}

\noindent In this section we explore the physical content of the KTCS's. The
probability of finding $l$ $(m,n)$ quanta of mode $a$ $(b,c)$ in state $%
\left| \xi ,p,q\right\rangle _{Kj}$ is given by

\begin{equation}
P_{lmn}(\xi ,p,q,K,j)=\left| _{Kj}\left\langle \xi ,p,q\right| \left.
l,m,n\right\rangle _{abc}\right| ^{2}=P_{n}(\xi ,p,q,K,j)\delta
_{m,n+p}\delta _{l,n+q}  \label{Plmn}
\end{equation}
with 
\begin{equation}
P_{n}(\xi ,p,q,K,j)=\frac{r^{2n}N_{Kj}^{2}(r^{2},p,q)I(\frac{n-j}{K})}{\rho
_{pq0}(n)}  \label{Pn}
\end{equation}
where $I(x)=1$ if $x$ is an integer and $I(x)=0$ if $x$ is a non-integer.
While the Kronecker symbols in (\ref{Plmn}) reveal the entanglement of the
modes in state $\left| \xi ,p,q\right\rangle _{Kj},$ the function $I(\frac{%
n-j}{K})$ in (\ref{Pn}) indicates that for any $K>1$ the number distribution
of state $\left| \xi ,p,q\right\rangle _{Kj}$ suffers an oscillation as
shown in Fig. 1 in which the TCS is also plotted that does not oscillate at
all. 
\begin{figure}[tbp]
\includegraphics[scale=0.45]{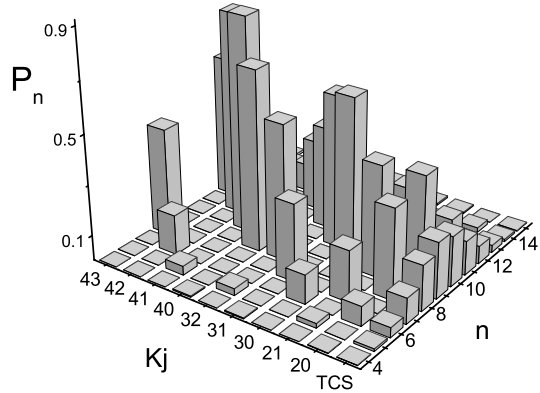}
\caption{3D bar plot showing oscillation of the number distribution, $%
P_{n}(\xi ,p,q,K,j),$ for $r =30, p=q=0$ and different combinations of $Kj$.
For comparison, also plotted is the case of TCS which does not oscillate at
all.}
\end{figure}
\begin{figure}[tbp]
\includegraphics[scale=0.3]{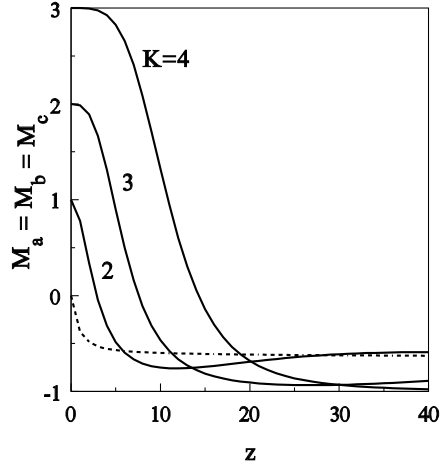}
\caption{Mandel parameters $M_{a}=M_{b}=M_{c}$ versus $z=r^{2}$ for $j=p=q=0$
and different $K$ indicated near the curve. For comparison, also plotted is
the case of TCS (dashed curve) which remains sub-poissonian in the whole
range of $z.$}
\end{figure}
\begin{figure}[tbp]
\includegraphics[scale=0.3]{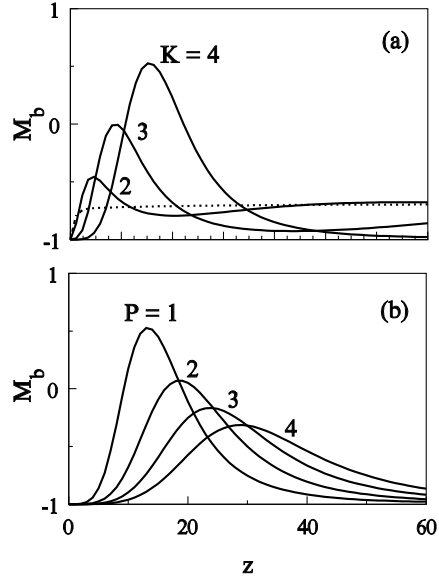}
\caption{Mandel parameter $M_{b}$ versus $z$ for $j=q=0$ while (a) $p=1 $
and $K=2,3,4;$ (b) $K=4$ and $p=1,2,3,4.$ The dashed curve in (a) is the
case of TCS which exhibits no maxima.}
\end{figure}
\begin{figure}[tbp]
\includegraphics[scale=0.3]{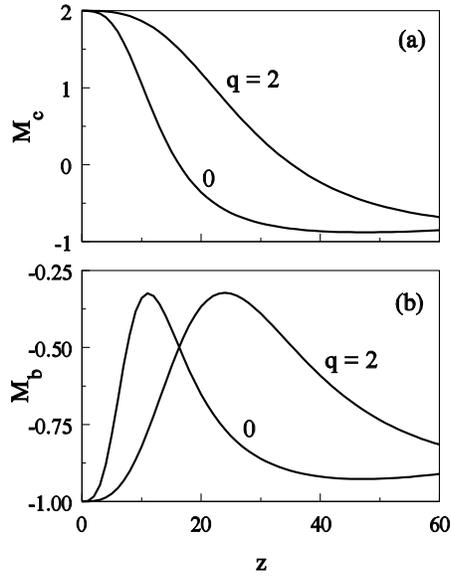}
\caption{Mandel parameters (a) $M_{c}$ and (b) $M_{b}$ versus $z$ for $K=3,$ 
$j=0,$ $p=2 $ while $q=0$ and $q=2.$}
\end{figure}
\begin{figure}[tbp]
\includegraphics[scale=0.3]{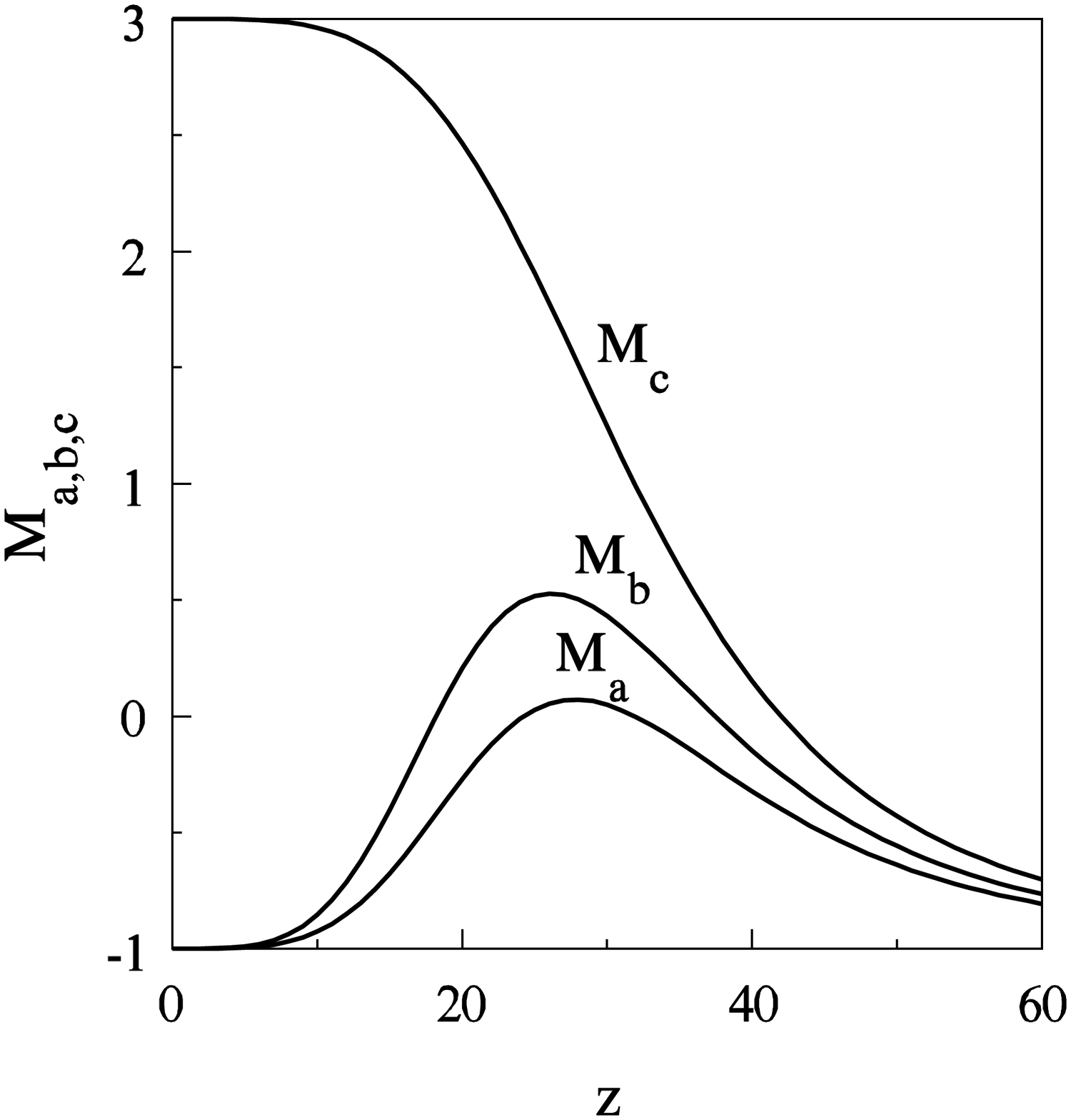}
\caption{Mandel parameters $M_{a},$ $M_{b}$ and $M_{c}$ versus $z$ for $K=4,$
$j=0,$ $p=1$ and $q=2.$}
\end{figure}
\begin{figure}[tbp]
\includegraphics[scale=0.3]{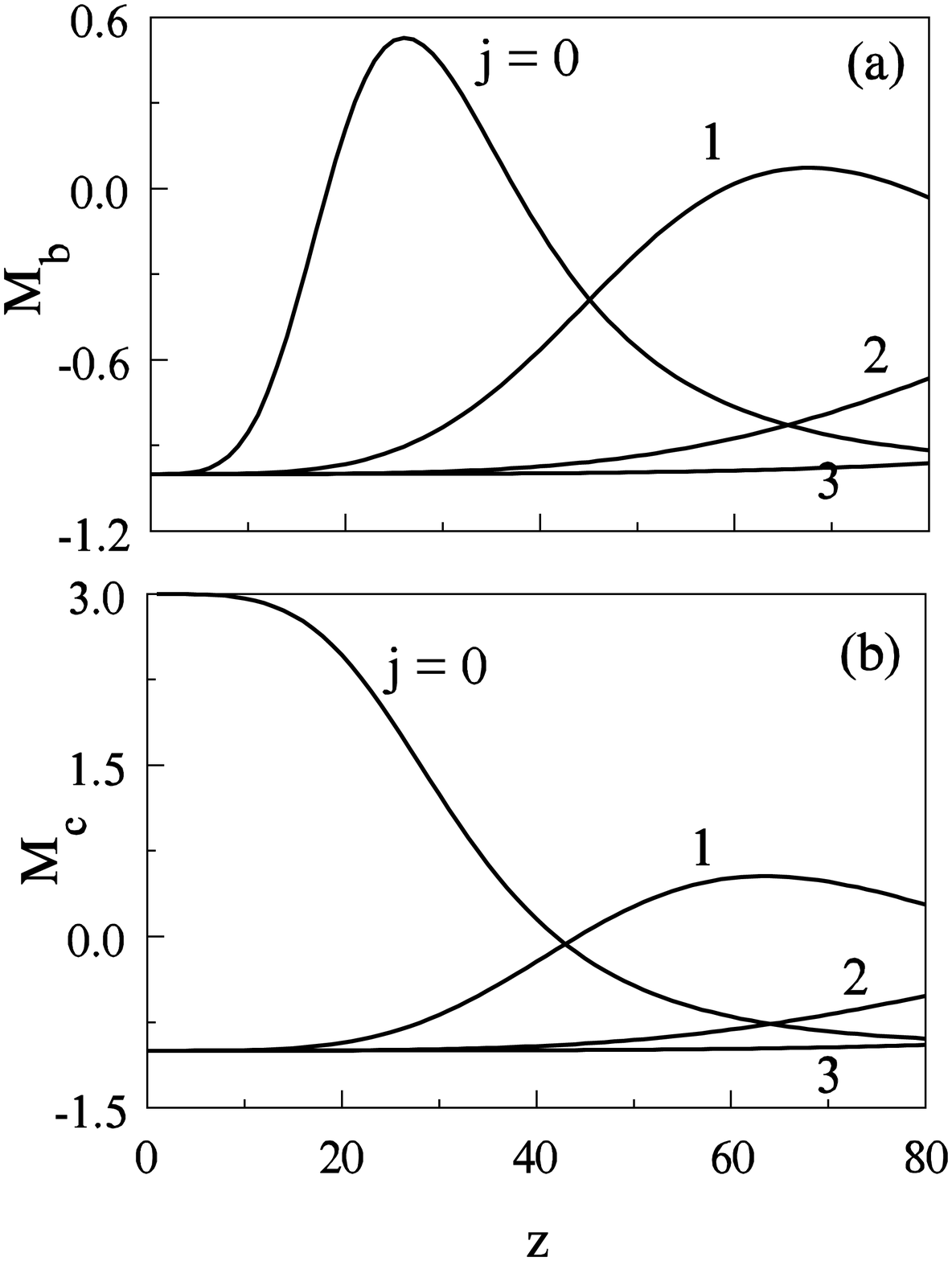}
\caption{Mandel parameters (a) $M_{b}$ and (b) $M_{c}$ versus $z$ for $K=4,$ 
$j=0,$ $1,$ $2,$ $3,$ $p=1$ and $q=2.$}
\end{figure}

Further information on inherent quantum statistics of state $\left| \xi
,p,q\right\rangle _{Kj}$ can be obtained from the Mandel parameter $M_{x}$
for mode $x$ \cite{mandel}

\begin{equation}
M_{x}=\frac{\left\langle \widehat{n}_{x}^{(2)}\right\rangle -\left\langle 
\widehat{n}_{x}\right\rangle ^{2}}{\left\langle \widehat{n}_{x}\right\rangle 
}  \label{Mx}
\end{equation}
where $\widehat{n}_{x}=\widehat{x}^{\dagger}\widehat{x}$ with $\widehat{x}$ the
annihilation operator of mode $x$ and the expectation value of the factorial
moment of the number operator $\left\langle \widehat{n}_{x}^{(l)}\right%
\rangle \equiv \left\langle \prod_{m=0}^{l-1}(\widehat{n}_{x}-m)\right%
\rangle $ is derived for our KTCS's in the form

\begin{equation}
\left\langle \widehat{n}_{a}^{(l)}\right\rangle =z^{l-q}N_{Kj}^{2}\frac{d^{l}%
}{dz^{l}}\left(\frac{z^{q}}{N_{Kj}^{2}}\right),  \label{nal}
\end{equation}

\begin{equation}
\left\langle \widehat{n}_{b}^{(l)}\right\rangle =z^{l-p}N_{Kj}^{2}\frac{d^{l}%
}{dz^{l}} \left(\frac{z^{p}}{N_{Kj}^{2}}\right),  \label{nbl}
\end{equation}

\begin{equation}
\left\langle \widehat{n}_{c}^{(l)}\right\rangle =z^{l}N_{Kj}^{2} \frac{d^{l}%
}{dz^{l}}\left(\frac{1}{N_{Kj}^{2}}\right),  \label{ncl}
\end{equation}
where $N_{Kj}\equiv N_{Kj}(z,p,q)$ and $z=r^{2}.$ An evident factor that
makes difference between modes is the charges $p,$ $q.$ For $p=q=0$ all the
three modes are ``identical". For $p>q=0$ $(q>p=0)$ mode $a$ $(b)$ and mode $%
c$ behave identically which are however distinct from mode $b$ $(a).$ For $%
p=q>0 $ modes $a$ and $b$ are similar while mode $c$ is dissimilar. Only
when $p\neq q$ and each of them acquires a positive integer the three modes
are well distinguishable one from another. Use of (\ref{nal}) in (\ref{Mx})
yields explicitly

\begin{equation}
M_{a}=\frac{2z^{2}\left( N_{Kj}N_{Kj}^{^{\prime \prime }}-N_{Kj}^{\prime
2}\right) +qN_{Kj}^{2}}{2zN_{Kj}N_{Kj}^{\prime }-qN_{Kj}^{2}},  \label{Ma}
\end{equation}

\begin{equation}
M_{b}=\frac{2z^{2}\left( N_{Kj}N_{Kj}^{^{\prime \prime }}-N_{Kj}^{\prime
2}\right) +pN_{Kj}^{2}}{2zN_{Kj}N_{Kj}^{\prime }-pN_{Kj}^{2}},  \label{Mb}
\end{equation}

\begin{equation}
M_{c}=\frac{z\left( N_{Kj}N_{Kj}^{^{\prime \prime }}-N_{Kj}^{\prime
2}\right) }{N_{Kj}N_{Kj}^{\prime }}  \label{Mc}
\end{equation}
where $N_{Kj}^{\prime }\equiv dN_{Kj}/dz$ and $N_{Kj}^{\prime \prime }\equiv
d^{2}N_{Kj}/dz^{2}.$

In Fig. 2 we plot $M_{c}$ as a function of $z$ for $j=p=q=0$ and different $%
K.$ Contrary to the TCS for which mode $c$ remains sub-poissonian (i.e. $%
M_{c}<0)$ in the whole range of $z,$ for KTCS's the mode is super-poissonian
(i.e. $M_{c}>0)$ at small values of $z$ but then becomes sub-poissonian when 
$z$ increases. The crossover point $z_{cross}$ at which the statistics
changes from super- to sub-poissonian moves to the right for greater values
of $K.$ At large values of $z$ the mode gets more antibunched in a higher
dimension.

For $p>q=0$ the shape of $M_{c}=M_{a}$ looks like that in Fig. 2 but for a
fixed $K$ the crossover point $z_{cross}$ shifts to the large-$z$ side with
increasing $p.$ For instance, when $K=3$ numerical calculations give $%
z_{cross}=7.5628,$ $12.0114,$ $16.3108$ and $20.5606$ for $p=0,$ $1,$ $2$
and $3,$ respectively. Concerning mode $b,$ its behavior is qualitatively
different as displayed in Fig. 3. Contrary to the TCS for which $M_{b}$
monotonically increases with $z,$ for KTCS's it exhibits a maximum which may
be located in the super-poissonian domain if the dimension $K$ is high
enough (e.g., if $K\geq 4$ when $p=1$ as illustrated in Fig. 3 (a)). The
effect of the charge $p$ is to pull down and right-shift the whole curve
(see Fig. 3 (b)) so that for a given $K$ the mode can be made entirely
sub-poissonian (if it was not so) by setting $p$ large enough (see, e.g.,
Fig. 3 (b) for $K=4:$ $M_{b}<0$ in the whole range of $z$ when $p\geq 3).$

For $p=q>0,$ as mentioned above, $M_{a}=M_{b}\neq M_{c}.$ Compared to the
situation $p>q=0$ the following properties hold. For mode $c$ we find the
relation $M_{c}(K,p=q>0)>M_{c}(K,p>q=0)$ in the whole range of $z$ except
for $z=0$ at which $M_{c}(K,p=q>0)=M_{c}(K,p>q=0)$, as seen from Fig. 4 (a).
For mode $b$ $(a)$ we find that $\max [M_{b}(K,p=q>0)]=\max [M_{b}(K,p>q=0)]$
but the maximum of $M_{b}(K,p>q=0)$ appears ``earlier'' than that of $%
M_{b}(K,p=q>0)$ when $z$ is increasing (see Fig. 4 (b)).

For $p>0,$ $q>0$ and $p\neq q$ each mode develops its own dependence on the
parameters and are well distinguishable as Fig. 5 shows.

The specific feature of KTCS's is their degeneracy as explained in Section
1. A given manifold characterized by a fixed $K>1$ consists of $K$
eigenstates enumerated by the parameter $j=0,$ $1$ $,$ $2,$ $...,$ $K-1.$
States with the same $K$ but different $j$ differ in their number
distribution which is dictated by the Kronecker symbols in Eq. (\ref{Plmn})
and by the function $I(\frac{n-j}{K})$ in Eq. (\ref{Pn}). These mean that
one cannot find $n$ quanta of mode $c$ $(b,a)$ in state $\left| \xi
,p,q\right\rangle _{Kj}$ if $n-j$ $(n-j-p,$ $n-j-q)$ is not a multiple of $K$
(see Fig. 1, for verification). States with the same $K$ but different $j$
differ also in their quantum statistics. So far (in Figs. 2 to 5) we have
treated only $j=0$ in which case, for arbitrary $K,$ $p$ and $q,$ we have $%
M_{c}(j=0,z\rightarrow 0)\rightarrow K-1$ and $M_{a,b}(j=0,z\rightarrow
0)\rightarrow -1.$ For $j>0,$ however, all the three modes are highly
antibunched at small values of $z$ independent of $K,$ $p$ and $q,$ i.e. $%
M_{a,b,c}(j>0,z\rightarrow 0)\rightarrow -1.$ Also, at small $z$ a state
with greater $j$ has higher degree of antibunching, i.e. its number
distribution profile is narrower. Those observations are demonstrated in
Fig. 6.

Another intriguing figure of merit is strong correlations between modes of
KTCS's. Such correlations are expected to be nonclassical. To verify it 
we examine the Cauchy-Schwarz inequality (CSI) \cite{cs} which describes a
classical correlation between two modes $x$ and $y$

\begin{equation}
J_{xy}=\left\langle \widehat{n}_{x}^{(2)}\right\rangle \left\langle \widehat{%
n}_{y}^{(2)}\right\rangle -\left\langle \widehat{n}_{x}\widehat{n}%
_{y}\right\rangle ^{2}\geq 0.  \label{CS}
\end{equation}
General expressions for expectation values of products of two factorial
moments $\widehat{n}_{x}^{(l)}\widehat{n}_{y}^{(m)}$ can also be
analytically derived for our KTCS's in terms of the normalization
coefficient (\ref{C}). As a result of derivation, we arrive at

\begin{equation}
\left\langle \widehat{n}_{a}^{(l)}\widehat{n}_{b}^{(m)}\right\rangle
=z^{m-p}N_{Kj}^{2}\frac{d^{m}}{dz^{m}}\left[ z^{l+p-q}\frac{d^{l}}{dz^{l}}%
\left( \frac{z^{q}}{N_{Kj}^{2}}\right) \right] ,  \label{ab}
\end{equation}

\begin{equation}
\left\langle \widehat{n}_{a}^{(l)}\widehat{n}_{c}^{(m)}\right\rangle
=z^{m}N_{Kj}^{2}\frac{d^{m}}{dz^{m}}\left[ z^{l-q}\frac{d^{l}}{dz^{l}}\left( 
\frac{z^{q}}{N_{Kj}^{2}}\right) \right] ,  \label{ac}
\end{equation}

\begin{equation}
\left\langle \widehat{n}_{b}^{(l)}\widehat{n}_{c}^{(m)}\right\rangle
=z^{l-p}N_{Kj}^{2}\frac{d^{l}}{dz^{l}}\left[ z^{m+p}\frac{d^{m}}{dz^{m}}%
\left( \frac{1}{N_{Kj}^{2}}\right) \right] .  \label{bc}
\end{equation}
Use of (\ref{nal}) and (\ref{ab}) - (\ref{bc}) in (\ref{CS}) yields
explicitly

\begin{eqnarray}
J_{ab} &=&N_{Kj}^{-3}\left\{ pq(1-p-q)N_{Kj}^{3}+24z^{3}N_{Kj}^{\prime
3}\right.  \nonumber \\
&&-2z^{2}N_{Kj}N_{Kj}^{\prime }\left[ \left( 2+7(p+q)-(p-q)^{2}\right)
N_{Kj}^{\prime }+4zN_{Kj}^{\prime \prime }\right]  \nonumber \\
&&+\left. 2zN_{Kj}^{2}\left[ 6pqN_{Kj}^{\prime }+z\left(
p+q-(p-q)^{2}\right) N_{Kj}^{\prime \prime }\right] \right\} ,  \label{Jab}
\end{eqnarray}

\begin{equation}
J_{ac} =2z^{2}N_{Kj}^{-3}\left\{ 12zN_{Kj}^{\prime
3}+q(1-q)N_{Kj}^{2}N_{Kj}^{\prime \prime } - N_{Kj}N_{Kj}^{\prime }\left[
\left( 2+7q-q^{2}\right) N_{Kj}^{\prime }+4zN_{Kj}^{\prime \prime }\right]
\right\} ,  \label{Jac}
\end{equation}

\begin{equation}
J_{bc}=2z^{2}N_{Kj}^{-3}\left\{ 12zN_{Kj}^{\prime
3}+p(1-p)N_{Kj}^{2}N_{Kj}^{\prime \prime }-N_{Kj}N_{Kj}^{\prime }\left[
\left( 2+7p-p^{2}\right) N_{Kj}^{\prime }+4zN_{Kj}^{\prime \prime }\right]
\right\} .  \label{Jbc}
\end{equation}
It is well-known that a nonclassical correlation violates the CSI \cite{c38a}, i.e. it
makes $J_{xy}<0.$ To assess degree of CSI violation we scale $J_{xy}$ to $%
\left\langle \widehat{n}_{x}\widehat{n}_{y}\right\rangle ^{2},$ i.e. we use
a quantity $G_{xy}$ defined by

\begin{equation}
G_{xy}=\frac{J_{xy}}{\left\langle \widehat{n}_{x}\widehat{n}%
_{y}\right\rangle ^{2}}  \label{Gxy}
\end{equation}
as a measure of CSI violation. For states with $j>0$ (i.e. $j=1,$ $2,$ $...,$
$K-1)$ we find that the CSI is always violated. However, for $j=0,$ though
the TCS also always violates the CSI, the KTCS's do not. The violation
depends on the type of correlation and the value of charges $p,$ $q.$ The
simulation reveals that $G_{ab}$ is always negative but $G_{ac}$ and $G_{bc}$
may be positive depending on the charges. More concretely, when both $p$ and 
$q$ are not greater than one we find that $G_{xy}<0$ $\forall x,y,K,z,$ i.e.
the CSI is fully violated. However, when $p$ $(q)=1$ and $q$ $(p)\geq 2$ the
quantity $G_{bc}$ $(G_{ac})$ remains always negative but $G_{ac}$ $(G_{bc})$
becomes positive at small values of $z$ and the higher the dimension $K$ the
wider the $z$-domain within which the CSI is not violated, i.e. there is a
partial violation of the CSI. Those behaviors are illustrated in Fig. 7.
Finally, when both $p\geq 2$ and $q\geq 2,$ both the quantities $G_{bc}$ and 
$G_{ac}$ violate the CSI only partially in the sense mentioned above. 
\begin{figure}[tbp]
\includegraphics[scale=0.3]{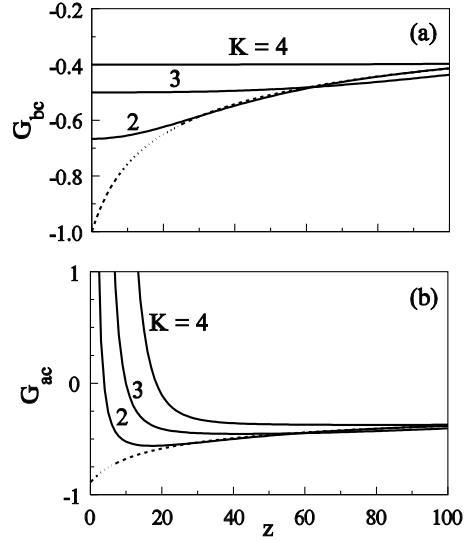}
\caption{Correlation measures (a) $G_{bc}$ and (b) $G_{ac}$ versus $z$ for $%
j=0,$ $p=1,$ $q=2$ and different $K$ as indicated near the curve. For
comparison, also plotted is the case of TCS (dashed curve) which remains
negative in the whole range of $z.$}
\end{figure}
\begin{figure}[tbp]
\includegraphics[scale=1.2]{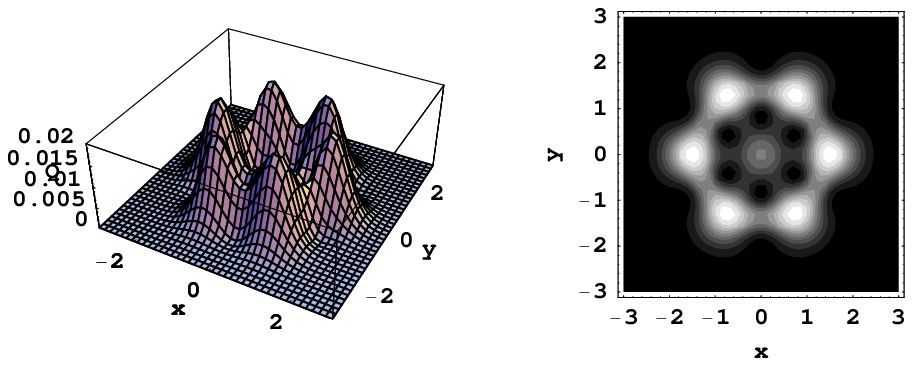}
\caption{Function $Q=\pi ^{3}Q_{\xi pq}^{Kj}(x,y)$ and its contour plot for $%
\xi =5,$ $p=q=0,$ $K=2$ and $j=0.$}
\end{figure}
\begin{figure}[tbp]
\includegraphics[scale=1.2]{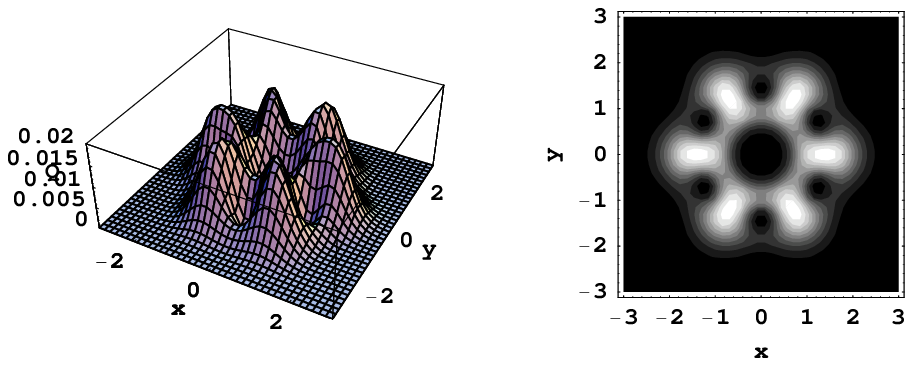}
\caption{Function $Q=\pi ^{3}Q_{\xi pq}^{Kj}(x,y)$ and its contour plot for $%
\xi =5,$ $p=q=0,$ $K=2$ and $j=1.$}
\end{figure}
\begin{figure}[tbp]
\includegraphics[scale=1.5]{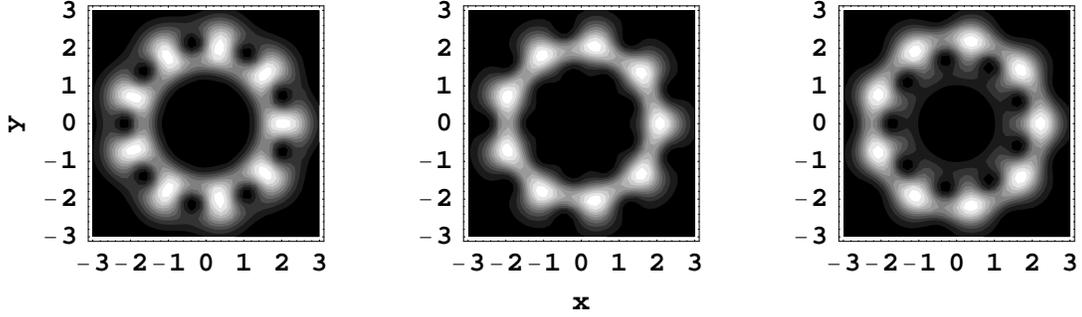}
\caption{Contour plots of $Q=\pi ^{3}Q_{\xi pq}^{Kj}(x,y)$ for $\xi =12,$ $%
p=q=0,$ $K=3$ with $j=0$ ( left), $1$ (middle) and $2$ (right).}
\end{figure}

We next study the phase-space characteristics of the KTCS's. For that
purpose we consider the three-mode Q-function defined as \cite{Q}

\begin{equation}
Q_{\xi pq}^{Kj}(\alpha ,\beta ,\gamma )=\frac{1}{\pi ^{3}}\left|
_{abc}(\alpha ,\beta ,\gamma \left| \xi ,p,q\right\rangle _{Kj}\right| ^{2}
\label{Q3}
\end{equation}
where $\alpha ,\beta ,\gamma \in \mathcal{C}$ and $|\alpha ,\beta ,\gamma
)_{abc}\equiv |\alpha )_{a}|\beta )_{b}|\gamma )_{c}$ with $|\alpha )_{a},$ $%
|\beta )_{b},$ $|\gamma )_{c}$ the usual coherent states (\ref{ucs}) of
modes $a,$ $b,$ $c.$ This function is non-negative definite, bounded and
normalized to unity

\begin{equation}
\int \int \int Q_{\xi pq}^{Kj}(\alpha ,\beta ,\gamma )d^{2}\alpha d^{2}\beta
d^{2}\gamma =1.
\end{equation}
Generally there are six variables associated with the real and imaginary
parts of $\alpha ,$ $\beta $ and $\gamma .$ For visualization let us confine
ourselves to a subspace determined by $\alpha =\beta =\gamma .$ In that
subspace the Q-function for KTCS's is calculated to be

\begin{equation}
Q_{\xi pq}^{Kj}(x,y)=\frac{1}{\pi ^{3}}N_{Kj}^{2}(\xi ,p,q)\text{ e}%
^{-3(x^{2}+y^{2})}(x^{2}+y^{2})^{p+q}\left| N_{Kj}^{-2}(\xi
(x-iy)^{3},p,q)\right| ^{2}  \label{Qxy}
\end{equation}
with $x=\mathop{\rm Re}(\alpha )$ and $y=\mathop{\rm Im}(\alpha ).$ We
represent in Figs. 8 and 9 the function $Q_{\xi pq}^{Kj}(x,y)$ and its
contour plot for $K=2,$ $j=0$ and $1.$ The figures clearly manifest
signatures of Schr\"{o}dinger-cat-like states: constructive (Fig. 8) and
destructive (Fig. 9) interference fringes between bell-like peaks (there are 
$3K$ bells for a given $K).$ The $K=2$ state with $j=0$ $(j=1)$ is called
even (odd) trio coherent state \cite{eoTCS1,eoTCS2}. We also present in Fig.
10 contour plots of the case of $K=3$ with $j=0,$ $1$ and $2$ which shows
nine interfering bells. Transparently the pronounced interference fringe
structures are $j$-dependent. It is the phase-space quantum interferences
between different TCS's (instead of being simply added) that bring about
copious nonclassical signatures of the KTCS. \vskip 0.5cm

\noindent \textbf{4. Physical realization}

\noindent After having studied physical properties of KTCS's we proceed to
find ways to realize them. In this section we are concerned with the context
of ion trap. Since ions can be trapped very efficiently and their
entanglement with the environment is extremely weak, trapped ions have
advantages for many purposes such as preparing various types of nonclassical
states (see e.g. \cite{n1,n2,n3,n4,n5,n6,n7}), simulating nonlinear
interactions \cite{ni}, demonstrating quantum phase transitions \cite{p1,p2}%
, establishing quantum search algorithms \cite{alg} and so on. The most
promising merit of trapped ion systems is perhaps the possibility to
implement scalable quantum computers \cite{qc} in which a number of ions are
involved \cite{few1,few2,few3}. Nevertheless, many tasks can still be done
even with a single ion. For instance, a controlled-NOT quantum logic gate
can be performed just by a single trapped ion \cite{s1,s2,s3,s4}: the target
qubit is stored in the ion internal electronic states while the external
quantized motional states, i.e. the phonon states, of the same ion serves as
the control qubit. Here we propose an experimental scheme to generate KTCS's
with $K=2$ in the vibronic motion of an ion which is trapped
in real three-dimensional $(3D)$ space. The situations corresponding to $1D$
and $2D$ were already considered in \cite{1d} and \cite{2d} for the single-
and two-mode case, respectively.

We first trap a two-level ion of mass $M$ by a $3D$ isotropic harmonic
potential characterized by the trap (phonon) frequency $\nu .$ Let $\widehat{%
a},$ $\widehat{b}$ and $\widehat{c}$ be the phonon annihilation operator in
the $x$-$,$ $y$- and $z$-axis, respectively. The ion is then simultaneously
driven by fourteen traveling-wave lasers (compare with the TCS case \cite{yi}%
) in the resolved sideband regime. The first thirteen lasers are all tuned
to be resonant with the sixth red motional sideband, i.e. their frequency is 
$\omega =\Delta -6\nu $ with $\Delta $ the energy difference between the two
electronic levels of the ion. For our purpose it is essential that the
lasers be judiciously configured. Namely, we shine the 1st (2nd, 3rd, 4th,
5th, 6th, 7th, 8th, 9th, 10th, 11th, 12th and 13th ) laser along the
direction connecting the coordinate origin $\{x,y,z\}=\{0,0,0\}$ to a point $%
\{x,y,z\}=\{1,1,1\}$ $(\{1,-1,1\},$ $\{1,1,-1\},$ $\{1,-1,-1\},$ $\{1,1,0\},$
$\{1,-1,0\},$ $\{1,0,1\},$ $\{1,0,-1\},$ $\{0,1,1\},$ $\{0,1,-1\},$ $%
\{1,0,0\},$ $\{0,1,0\}$ and $\{0,0,1\}).$ As for the 14th laser, it must be
in resonance with the electronic transition, i.e. its frequency is equal to $%
\Delta ,$ but its propagation direction is unimportant. The Hamiltonian of
the ion-phonon-laser system is $(\hbar =1)$

\begin{equation}
H=\frac{\Delta }{2}\sigma _{z}+\nu (\widehat{a}^{\dagger}\widehat{a}+\widehat{b}%
^{\dagger}\widehat{b}+\widehat{c}^{\dagger}\widehat{c})+H_{int}  \label{H}
\end{equation}
where

\begin{equation}
H_{int}=\sum_{l=1}^{14}\left[ \Omega _{l}\exp [-i(\omega _{l}t+\varphi
_{l})+i\mathbf{k}_{l}\widehat{\mathbf{R}}_{l}]\sigma _{+}+\text{H.c.}\right] 
\label{Hint}
\end{equation}
with $\Omega _{l}$ the Rabi frequencies, $\omega _{1,2,...,13}=\omega ,$ $%
\omega _{14}=\Delta ,$ $\varphi _{l}$ the phases, $\widehat{\mathbf{R}}_{l}$
the position operator along the laser direction determined by the wave
vector $\mathbf{k}_{l},$ $\sigma _{z}=\left| e\right\rangle \left\langle
e\right| -\left| g\right\rangle \left\langle g\right| ,$ $\sigma _{+}=\sigma
_{-}^{\dagger}=\left| e\right\rangle \left\langle g\right| $ and $\left|
g\right\rangle $ $(\left| e\right\rangle )$ the ion's electronic ground
(excited) state. Assuming $k_{l}=k$ $\forall l$ there is a single Lamb-Dicke
parameter $\eta =k/\sqrt{2M\nu }$ which measures the localization of the ion
relative to the laser wavelength. In terms of $\eta $ we obtain, in an
interaction picture with respect to $H_{0}=H-H_{int}$, the interaction
Hamiltonian of the form 
\begin{equation}
\mathcal{H}_{int}=\text{e}^{-\eta ^{2}/2}\sum_{m=0}^{\infty }\frac{(-\eta
^{2})^{m}}{m!}\left[ -\eta ^{6}\sum_{l=1}^{13}\Omega _{l}\text{e}^{-i\varphi
_{l}}\frac{\left( \widehat{A}_{l}^{\dagger}\right) ^{m}\widehat{A}_{l}^{m+6}}{%
(m+6)!}+\Omega _{14}\text{e}^{-i\varphi _{14}}\frac{\left( \widehat{A}%
_{14}^{\dagger}\right) ^{m}\widehat{A}_{14}^{m}}{m!}\right] \sigma _{+}+\text{ H.c.%
}  \label{hint}
\end{equation}
In the Lamb-Dicke limit $\eta \ll 1$ we can retain only the $m=0$ term in (%
\ref{hint}) and reduce it to

\begin{equation}
\mathcal{H}_{int}=\left[ -\frac{\eta ^{6}}{6!}\sum_{l=1}^{13}\Omega _{l}%
\text{e}^{-i\varphi _{l}}\widehat{A}_{l}^{6}+\Omega _{14}\text{e}^{-i\varphi
_{14}}\right] \sigma _{+}+\text{H.c.}  \label{hh}
\end{equation}
Thanks to our configuration for the lasers the operators $\widehat{A}%
_{1,2,...,13}$ are expressed through $\widehat{a},$ $\widehat{b}$ and $%
\widehat{c}$ as 
\begin{equation}
\widehat{A}_{1,2}=\widehat{a}\pm \widehat{b}+\widehat{c},\text{ }\widehat{A}%
_{3,4}=\widehat{a}\pm \widehat{b}-\widehat{c},  \label{A0}
\end{equation}
\begin{equation}
\widehat{A}_{5,6}=\widehat{a}\pm \widehat{b},\text{ }\widehat{A}_{7,8}=%
\widehat{a}\pm \widehat{c},\text{ }\widehat{A}_{9,10}=\widehat{b}\pm 
\widehat{c},  \label{A1}
\end{equation}
\begin{equation}
\widehat{A}_{11}=\widehat{a},\text{ }\widehat{A}_{12}=\widehat{b},\text{ }%
\widehat{A}_{13}=\widehat{c}.  \label{A2}
\end{equation}
If we control the laser intensities and phases in such a way that

\begin{equation}
\Omega _{1,2,3,4}=\frac{1}{2}\Omega _{5,6,7,8,9,10}=\frac{1}{4}\Omega
_{11,12,13}=\Omega ,  \label{omega}
\end{equation}

\begin{equation}
\varphi _{1,2,3,4}=\varphi _{11,12,13}=\varphi +\pi ,\text{ }\varphi
_{5,6,7,8,9,10}=\varphi ,\text{ }\varphi _{14}=\pi ,  \label{phai}
\end{equation}
then use of Eqs. (\ref{A0}) - (\ref{A2}) in (\ref{hh}) yields 
\begin{equation}
\mathcal{H}_{int}=\zeta \left[ (\widehat{a}\widehat{b}\widehat{c})^{2}-\xi
^{2}\right] \sigma _{+}+\text{H.c.},  \label{hhh}
\end{equation}
thanks to the identity
\begin{equation}
\sum_{l=1}^4 \widehat{A}_{l}^6-2\sum_{m=5}^{10} \widehat{A}_{m}^6+4\sum_{n=11}^{13} \widehat{A}_{n}^6
\equiv 360 (\widehat{a}\widehat{b}\widehat{c})^2.
 \label{identity}
\end{equation}
The Hamiltonian (\ref{hhh}) is the central result for the physical realization under consideration. 
The new parameters $\zeta ,$ $\xi $ appearing in (\ref{hhh}) are
controllable and given simply by

\begin{equation}
\zeta =\frac{\Omega \eta ^{6}}{2}\exp (-i\varphi ),  \label{zeta}
\end{equation}

\begin{equation}
\xi ^{2}=\frac{2\Omega _{14}}{\Omega \eta ^{6}}\exp (i\varphi ).  \label{xi}
\end{equation}
Since the trapped ion is well isolated the dominant channel 
of decoherence is via the spontaneous decay. Then the time evolution 
of the system density operator $\rho $
is governed by the following master equation 
\begin{equation}
\frac{d\rho }{dt}=-i[\mathcal{H}_{int},\rho ]+\Gamma \left( \sigma _{-}\rho
\sigma _{+}-\frac{\sigma _{+}\sigma _{-}\rho }{2}-\frac{\rho \sigma
_{+}\sigma _{-}}{2}\right)  \label{rr}
\end{equation}
where $\Gamma $ accounts for the spontaneous decay rate of the ion being 
in its electronic excited state $\left| e\right>$.

\begin{figure}[tbp]
\includegraphics[scale=0.9]{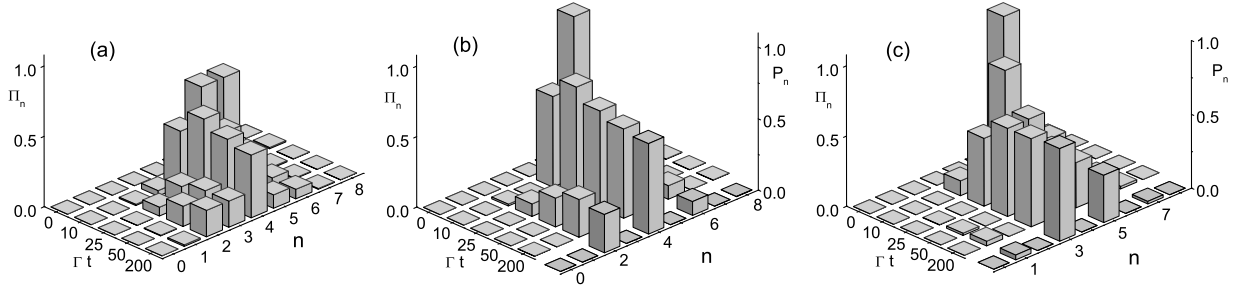}
\caption{3D bar plots of phonon number distribution $\Pi _{n}$ at different
values of $\Gamma t$ for $\xi =8$, $p=q=0$, $\zeta =0.02$ and $l=3$. (a) $%
w=0.5$: no KTCS's are generated since in the long-time limit $\Pi _{n}$ 
is non-zero at both even and odd $n$, i.e. it does not display 
required oscillation. (b) $w=0$: generation of state $\left|
\xi ,p,q\right\rangle _{20}$ is justified in the long-time limit by coincidence 
of $\Pi _{n}$, Eq. (\ref{Pin}), with $P_{n}\equiv P_{n}(\xi ,p,q,2,0)$, Eq. (\ref{Pn}).  (c) $w=1$:
generation of state $\left| \xi ,p,q\right\rangle _{21}$ is justified in the long-time limit 
by coincidence of $\Pi _{n}$, Eq. (\ref{Pin}), with $P_{n}\equiv P_{n}(\xi ,p,q.2,1)$, Eq. (\ref{Pn}).}
\end{figure}

It is clear that the system ceases to evolve when the ion's fluorescence stops.
Hence, the ``dark'' stationary solution of Eq. (\ref{rr}) in the long-time
limit has the ansatz

\begin{equation}
\rho (\infty )=\left| g\right\rangle \left| \Psi \right\rangle
_{xyz}\left\langle \Psi \right| \left\langle g\right|   \label{ansatz}
\end{equation}
where $\left| \Psi \right\rangle _{xyz}$ solely determines the phonon state.
Setting $d\rho (\infty )/dt=0$ in Eq. (\ref{rr}) and using properties of the
operators $\sigma _{\pm }$ we arrive at an equation for $\left| \Psi
\right\rangle _{xyz},$

\begin{equation}
(\widehat{a}\widehat{b}\widehat{c})^{2}\left| \Psi \right\rangle _{xyz}=\xi
^{2}\left| \Psi \right\rangle _{xyz},  \label{con}
\end{equation}
where $\xi ^{2}$ depends on the experiment parameters via (\ref{xi}).

For $\left| \Psi \right\rangle _{xyz}$ to be a desired KTCS $\left| \xi
,p,q\right\rangle _{20}$ or $\left| \xi ,p,q\right\rangle _{21}$ we need to
prepare an appropriate initial state. Furthermore, the generation time is
influenced by all the parameters involved. To see these let us simulate Eq. (%
\ref{rr}) by the Monte Carlo Wave-Function approach \cite{mcwf} with an
initial state of the form 

\begin{equation}
\left| \Phi _{w}(0)\right\rangle =\left| e\right\rangle \left( \sqrt{w}%
\left| \Psi _{l0}\right\rangle _{xyz}+\sqrt{1-w}\left| \Psi
_{l1}\right\rangle _{xyz}\right) 
\end{equation}
where $0\leq w\leq 1$ and 

\begin{equation}
\left| \Psi _{lj}\right\rangle _{xyz}=\left| 2l+q+j\right\rangle _{x}\left|
2l+p+j\right\rangle _{y}\left| 2l+j\right\rangle _{z}
\end{equation}
with a non-negative integer $l$ and $j=0$ or $1.$ Since the dynamics under
consideration does not mix the parity, at time $t>0$ we have 

\begin{equation}
\left| \Phi _{w}(t)\right\rangle =\sum_{m=0}^{\infty }\left( G_{m}(t)\left|
g\right\rangle +E_{m}(t)\left| e\right\rangle \right) \left( \sqrt{w}\left|
\Psi _{m0}\right\rangle _{xyz}+\sqrt{1-w}\left| \Psi _{m1}\right\rangle
_{xyz}\right) 
\end{equation}
where $G_{m}$ and $E_{m}$ are some time-dependent coefficients to be
simulated. In Fig. 11 we plot the phonon number distribution 

\begin{equation}
\Pi _{n}(t)=\left| \left\langle \Phi _{w}(t)\right| \left. n+q\right\rangle
_{x}\left| n+p\right\rangle _{y}\left| n\right\rangle _{z}\right| ^{2}
\label{Pin}
\end{equation}
at different times. Obviously, for $0<w<1,$ $\Pi _{n}$ may be nonzero at any 
$n.$ In particular, for $w=0.5$ there is no oscillation at all in $\Pi _{n}$
in the long-time limit (see Fig. 11 (a)). Thus, a desired KTCS $\left| \xi
,p,q\right\rangle _{20}$ or $\left| \xi ,p,q\right\rangle _{21}$ results
only when either $w=0$ or $w=1$ as seen from Fig. 11 (b) or Fig. 11 (c).
These figures show that in the course of time the phonon number distribution 
$\Pi _{n}(t)$ changes gradually and finally tends to coincide with $%
P_{n}(\xi ,p,q,2,0)$ if $w=0$ (Fig. 11 (b)) or with $P_{n}(\xi ,p,q,2,1)$ if 
$w=1$ (Fig. 11 (c)), implying successful generation of the target state $%
\left| \xi ,p,q\right\rangle _{20}$ or $\left| \xi ,p,q\right\rangle _{21}.$

\begin{figure}[tbp]
\includegraphics[scale=0.9]{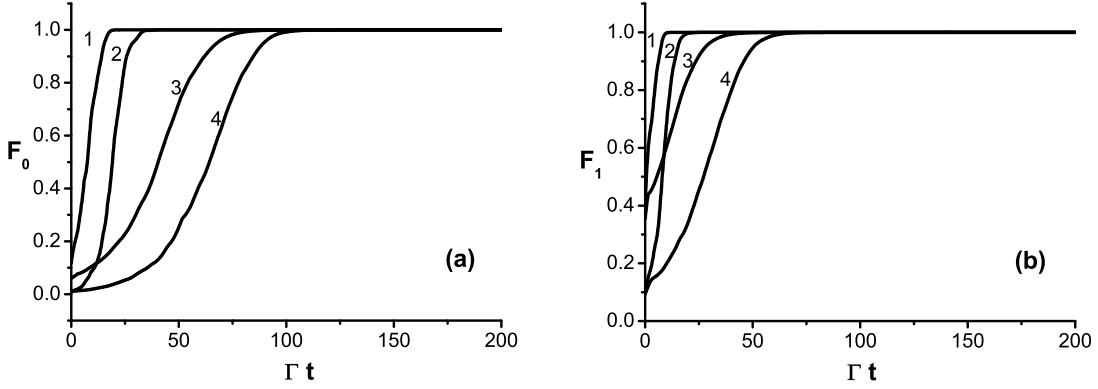}
\caption{Fidelities (a) $F_{0}$ and (b) $F_{1}$ as a function of $\Gamma t$
for different sets of parameters: Curves 1, 2, 3 and 4 correspond to 
$\{\xi, \zeta/\Gamma, p=q, l\}=\{10,0.005,2,0\},$ $\{10,0.005,0,0\},$ 
$\{6,0.005,0,0\}$ and  $\{10,0.002,0,0\},$ respectively.} 
\end{figure} 

To assess quality of the resulting state as well as the generation time of
our scheme we examine time-dependence of the fidelity 

\begin{equation}
F_{j}(t)=|\left\langle \Phi _{j}(t)\right| \left. \xi ,p,q\right\rangle
_{2j}|^{2}.
\end{equation}
In Fig. 12 we represent $F_{j}$ as a function of $\Gamma t$ for various sets
of parameters. Although initially the transient behaviors happen differently for
different parameters' sets, after some period of time a stationary regime is inevitably 
established in all cases. It also follows from the figure that the generation time, 
i.e. the time needed to reach the stationary 
regime, is shorter for greater $p,q$ (compare Curves 1 and 2),  $\xi $ (compare Curves 2 and 3) 
as well as $\zeta$ (compare Curves 2 and 4). 
For a wide range of parameters we find out that $%
1-F_{j}$ becomes less than $10^{-3}$ for $\Gamma t\geq 200.$ The desired
KTCS is thus produced with high purity within a relatively short period of time
(compared, e.g.,  with the TCS generation time \cite{yi}). The KTCS obtained in our
scheme is also stable since its appearance is accompanied by the ``dark''
ion which is to be found in the ground state $\left| g\right\rangle $ and therefore does not 
interact with the laser fields any longer. \vskip 0.5cm

\noindent \textbf{5. Conclusion}

\noindent We have introduced and studied KTCS's which are both higher-order
and multimode. Being superposed of phase-correlated TCS's, a KTCS is indeed a
new physical state as dictated by the quantum mechanics superposition
principle. The novel nonclassical features of KTCS's, as compared to TCS's,
were demonstrated in detail and an experimental generation scheme for $K=2$
was also presented in the ion trap context. Since these KTCS's are 
entangled states, they do promise potential implementations in quantum
information processing and quantum computation. In particular, their
three-mode nature would be crucial for tasks such as quantum controlled
teleportation or/and quantum telecloning of certain types of
continuous-variable states as well as inequality-free ``one-shot'' tests of
local hidden-variable theories, etc. Such applications of KTCS's are being
under way and will be reported elsewhere. \vskip 0.5cm

\noindent \textbf{Acknowledgments}

\noindent The authors thank the KIAS Quantum Information Group for useful
discussions. H.S.Y. was supported by R\&D Program for Fusion Strategy of
Advanced Technologies MI-0326-0830002, B.A.N. by KIAS R\&D Fund No.
03-0149-002 and J.K. by Korea Research Foundation Grant No.
KRF-2002-070-C00029.

\end{document}